\documentclass[epj,nopacs, referee]{svjour}
\usepackage{latexsym}
\usepackage{graphics}
\def\be{\begin{equation}}
\def\ee{\end{equation}}
\def\bea{\begin{eqnarray}}
\def\eea{\end{eqnarray}}
\newcommand{\il}{~}

\usepackage[centertags]{amsmath}
\usepackage{caption}
\usepackage{ctable}
\usepackage{tabularx}
\usepackage{color}
\usepackage{amssymb}
\newcommand{\ti}[1]{\mbox{\tiny{#1}}}
\newcommand{\api}[1]{\ ^{\mbox{\tiny{(#1)}}}\! }%

\begin{document}

\title{On the electromagnetic
emission from charged test particles  in a five dimensional
spacetime}
\author{D. Pugliese\inst{1,2}, G. Montani\inst{1,3,4}, V. Lacquaniti\inst{5}}
\offprints {G. Montani}
\mail{giovanni.montani@frascati.enea.it}
\institute{
Dipartimento di Fisica, Universita' di  Rome, ``Sapienza'', - Piazzale Aldo
Moro 5, 00185 Roma, Italy \and School of Mathematical Sciences, Queen Mary, University of London,
Mile End Road, London E1 4NS, United Kingdom.  \email{d.pugliese.physics@gmail.com}
\and
ENEA-C. R. Frascati U.T.
Fus. (Fus. Mag. Lab.), via E.Fermi 45, I-00044, Frascati, Roma, Italy \email{giovanni.montani@frascati.enea.it}
\and
INFN-(Istituto Nazionale di Fisica Nucleare)-,  Sezione  Roma1 - Piazzale Aldo
Moro 5, 00185 Roma, Italy
\and
Physics Department ``E.Amaldi'', University of Rome, ``Roma Tre"',
Via della Vasca Navale 84, I-00146, Rome, Italy \email{lacquaniti@fis.uniroma3.it}}

\date{Received: date / Revised version: date}

\abstract{
We study   the motion  of charged particles radially falling in a
class of static and electromagnetic-free,  five-dimensional
Kaluza-Klein backgrounds. Particle dynamics in such spacetimes is
explored by an approach \emph{a l$\acute{a}$}   Papapetrou.  The
electromagnetic radiation   emitted by these particles is studied, outlining the new features emerging in the spectra for the five-dimensional case. A
comparison with the dynamics in the four dimensional
counterpart, \textit{i.e} the Schwarzschild background, is performed.\keywords{Kaluza Klein theory--Generalized Schwarzschild solution (GSS)--Test particle orbits--Emission spectra}}

\titlerunning{On the electromagnetic
emission from charged test particles  in a five dimensional spacetime}
\authorrunning{D. Pugliese et al.}
 \maketitle

\section{Introduction}\label{secintro}
Multidimensional theories are considered a great promise of the
present physics as possible candidates for the grand unification
theory. Great attention  is therefore devoted to any  available test
to probe the validity of these theories (see for example \cite{Matsuno:2011ca,Rizzo:1999qb,Eingorn:2010wi,Moutsopoulos:2011ez,Stelea:2009ur,Bonnevier:2011km,Okawa:2011fv,Tomizawa:2011mc}
 and also \cite{Moon:2011sz,Becar:2011fc,Yamada:2011br,Inte}).
The five-dimensional (5D) Kaluza-Klein (KK) theory unifies
electromagnetism and gravitation in a geometrical picture,
reformulating  the internal U(1) gauge symmetry in a geometrical one
by the introduction of a  fifth dimension. Translations in this
space are a generalization of  the gauge transformation (see for
instance \cite{Librogra,PSWesson,P.D.B.Collins,Cianfrani:2009wj}).
    {In this work we consider a 5D-background, known as Generalized
Schwarzschild Solution
(GSS) \cite{Overduin:1998pn,Wesson:1999nq},  solution of the
electromagnetic-free KK-equations. }
We investigate the motion of  charged particles radially falling
in GSS spacetimes. Electromagnetic (EM) radiation emitted by such
particles is studied as a perturbation of the background metric.
Spectral observations could be used to test the validity of a
multidimensional theory: the presence of an additional dimension
could be  recognized  via the  deformation induced on the spectra
predicted  by the General Relativity (see for example \cite{Kokkotas:1999bd}).
The paper is organized as  follows; in Sec.~\ref{sec:GSSMetric} we
discuss some general aspects of the GSS, in Sec.~\ref{sec:Papapetroumotion} we briefly review some fundamental
statements of the KK-particle dynamics, in
Sec.~\ref{sec:perturbation} EM-perturbations in  GSS spacetimes are
analyzed. Finally in Sec.~\ref{sec:concldisc} concluding remarks
follow.
\section{Generalized Schwarzschild Solution}\label{sec:GSSMetric}
The GSS is a family   of free-electromagnetic, static 5D-vacuum
solutions of KK-equations. Adopting  spherical
polar coordinates the GSS reads\footnote{With latin capital letters $A$ we
label the five-dimensional indices, they run in $\{0,1,2,3,5\}$,
Greek  indices $\alpha$ run from 0 to 3,  $x^{5}$ is the angle
parameter for the fifth circular dimension. We consider metric of
$\{+,-,-,-,-\}$ signature. Here $c=G=1$.}
\begin{equation}\label{CGMriunone}
 ds_{\ti{(5)}}^{2}=\Delta^{\epsilon
k}dt^{2}-\Delta^{-\epsilon (k-1)}dr^{2}-r^{2}\Delta^{1-\epsilon
(k-1)}d\Omega^{2}-\Delta^{-\epsilon }(dx^{5})^2
\end{equation}
where $d\Omega^2\equiv\left(d\theta^2+\sin^2\theta d\varphi^2\right)$,   $\Delta\equiv\left(1-2M/r\right)$, $M>0$ is a constant,
$\epsilon=\left(k^{2}-k+1\right)^{-1/2}$ and $k$ is a positive
dimensionless parameter. On the 4D-counterpart metric (the spacetime
section $dx^{5}=0$), the Schwarzschild solution is recovered for
$k\rightarrow\infty$, where in this limit $M$ is the Schwarzschild
mass.
The region $k\geq0$ and $\epsilon\geq0$ of the metric parameter is generally taken into account to investigate the
physical properties of
solutions. In  \cite{Overduin:1998pn},
for example, is showed how positive
density of solution requires $k>0$ and  for
positive mass (as measured at infinity) one must have $\epsilon k >
0$.
For a
review about the  constraint on $k$ parameter see  \cite{Overduin:1998pn} and  \cite{Lacquaniti:2009wc}.
To test the
validity of   GSS model, the predictions elaborated within a GSS scenario  have been compared with the observed data. Each comparison
gives a peculiar estimation for $k$; all these different estimations
are based on different tests assumed to prove the model validity.
For example,  a particular value of $k$, adapted to
fit the prediction of the standard gravity tests with experimental
data, has been associated to the Sun
\cite{Overduin:1998pn,Xu:2007dc}.
In \cite{Overduin:2000gr}, experimental constraints on equivalence
principle violation in the solar system translate in a $k>5.\times
10^7$. We could conclude therefore that extra dimensions  play thus a negligible role in the solar
system dynamics. Meanwhile, by measures of the surface gravitational
potential in \cite{PoncedeLeon:2006xs,PoncedeLeon:2007bm}, the Sun
seems to be characterized by a $k=2.12$. On the other hand  all the
standard tests on light-bending around the Sun, or the perihelion
precession of Mercury, constrain $k \gtrsim 14$.

{Nevertheless we must note here that the metric parameter $k$
 characterizes the spacetime external to any
astrophysical object described by the selected GSS; the Sun is therefore only one particular astrophysical source to be investigated, but there
is one different value of the $k$ parameter associated to one
different astrophysical source. }

GSS metrics are asymptotically flat. For $r\rightarrow 2M$ from the
right, $g_{tt}\rightarrow0$ and $g_{55}\rightarrow\infty$, therefore
the  length of the extra dimension increases as well as $r$
approaches  $2M$. The
5D-Kretschmann scalar and the square of the 4D-Ricci tensor are
divergent in such a  point.
At first order in $(1/k)$ for large $k$, namely the
quasi-Schwarzschild regime, the 5D-Kretschmann scalar diverges in
$r=2M$ while the square of the 4D-Ricci tensor is identically zero.
These solutions represent extended  objects. The induced
scalar matter has a  trace-free energy-momentum tensor
$T_{\nu}^{\mu}$ and its  gravitational mass
$M_g=\int(T^0_0-T^1_1-T^2_2-T^3_3)\sqrt{-g_4}dV_3$  is $M_g=\epsilon
kM$ at infinity and  goes to zero at $r=2M$, for every $k$, where
$g_4$ is the determinant of the 4D-metric and $dV_3$ is the ordinary
spatial 3D-volume element. In the Schwarzschild limit however,
$M_g=M$ for every $r$, therefore also in $r=2M$.

Hence, for every finite values of $k$, GSS are  naked
singularities surrounded by the induced scalar matter, and only in the limit
$k\rightarrow\infty$ they are black holes with an horizon in $r=2M$
and vacuum for $r>2M$.
In this work we consider GSS with  $k>1$ as  the (5D-vacuum) spacetime, surrounding a compact object with a radius
$R>2M$.

A solution of
KK-equations  \cite{Lacquaniti:2009rq}, in the region $r\leq R$, that matches the GSS in $r=R$
was obtained in \cite{KKStars}, by a numerical integration, considering the energy
momentum tensor of a 4D-perfect fluid with an  equation of
state $p=C\rho^b$, where $p$ is the pressure, $\rho$ the density, and the constants ${C,b}$ are real numbers. The existence of such configuration allows to remove the naked singularity providing physical significance to our analysis.
A solution of the form $ds_{\ti{(5)}}^{2}=\tilde{f} dt^{2}-\tilde{g} dr^{2}-\tilde{h} d\Omega^{2}-\tilde{q} (dx^{5})^2$ is considered, where  $(\tilde{f}, \tilde{g}, \tilde{h}, \tilde{q})$ are functions of the $r$ only (see also \cite{Capozziello:2011nr} for an analogue analysis on the hydrostatic equilibrium and stellar structure in in $f(R)$ theories.).
For a discussion on the Papapetrou approach to matter and geometry and the properties of the interior solutions we refer to \cite{Lacquaniti:2009rq,KKStars}.

\begin{figure}
\resizebox{.51\textwidth}{!}{%
\includegraphics[scale=.71]{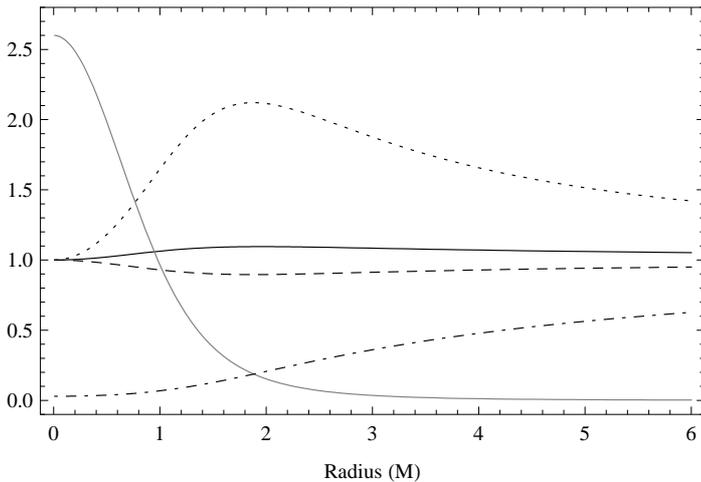}}
\caption{Metric coefficients $\tilde{f}$ (dotdashed line), $\tilde{g}$ (dotted line), $\tilde{h}$ (dashed line),  $\tilde{q}$ (black line) and the density $\rho$ (gray line) as functions of the radial coordinate in unit of mass $M$, where $p=C \rho^b$. Here $k=5$, $C=2/5$ and $b=1.329$. }
\label{Fig:stellaRoma}
\end{figure}

In Fig.\ref{Fig:stellaRoma}. the  solution with $C=2/5$, $b=1.329$ and $k=5$ is plotted. Metric components $(\tilde{f}, \tilde{g}, \tilde{h}, \tilde{q})$ and the functions $(p, \rho)$ are plotted as function of the radial coordinate $r$.
The radius $R$ of the configuration has been numerically evaluated via the condition $\rho(R)=0$.

We consider the  particle dynamics and the fields propagating   in a
region $r>R\equiv2(1+10^{-a})M$, with $a>0$ constant. We compare the
dynamic in GSS with the dynamic  in the Schwarzschild geometry
(See \cite{Casadio:2003iv} for a similar problem applied to the Janis-Newman-Winicour spacetime).
\section{Papapetrou approach to particle dynamics}\label{sec:Papapetroumotion}

The physical characterization of the considered
solution as a real four-dimensional spacetime
requires that the process of dimensional reduction
be satisfactory addressed.

The fundamental requirement to make unobservable
the fifth dimension is the existence of a closed
topology on the extra-dimension,
which allows to compactify the spacetime,
reducing it to be physically undistinguishable
from a four-dimensional one.

The basic request to deal with a closed
topology is the periodicity of the metric field on
the fifth coordinate, which allows to expand the
Einstein-Hilbert action in  Fourier series and
so to develop the dynamics of the different harmonics.
This approach is intrinsically different from
the cylindricity condition, due to Kaluza,
which prescribes the independence of the metric tensor
of the fifth coordinate, directly within the field
equations. However, if on one hand we truncate the
Einstein-Hilbert action to the zero-order and,
on the other hand, we impose by hands the
closure of the fifth coordinate in the field
equations independent of $x^5$, the two
approaches formally overlap.
In fact, the variation of the zero-order Lagrangian
(after the integration on the extra-coordinate
has been performed) coincides (without
further restrictions) with the Einstein equations
in which the four-dimensional quantities are
explicitly outlined. It is just this coincidence
of the two approaches in the specified limit which
gives to the five-dimensional KK-model
\cite{Cianfrani:2005em} a privileged
character among the other numbers of extra-dimensions
(where a non-trivial integration on the extra-dimensions
is also required on the field equations too
\cite{Cianfrani:2009wj,Cianfrani:2006sc,Montani:2009wg}).

In the case of the GSS, the situation is further simplified
by the absence of an electromagnetic potential
$g_{5\mu} = 0$, which makes diagonal the metric
tensor, reduced to provide a scale factor for
each space coordinate. In particular the
scalar field $\phi$, which governs the evolution
of the fifth dimension, can be regarded as matter
in both the two points of view elucidated above,
just in view of the possibility to split
the five-dimensional Einstein equations into
a four-dimensional set, characterized by
a scalar matter source.

Our approach to include matter in the vacuum KK-theory, is based on the analysis in \cite{Lacquaniti:2009yy,Lacquaniti:2009cr,Lacquaniti:2009rh},
which relays on the idea that, since the fifth dimension
is compactified to planckian-like scales,
it makes no sense to handle with a classical
fifth component of the particle velocity and therefore
the only viable treatment of matter sources is
through the energy-momentum description. Such a
point of view leads to adopt a Papapetrou scheme
\cite{papapetrou} to fix the dimensionally reduced
equation of motions for fields and macroscopic matter.

Since the Papapetrou procedure is based on the use
of the field equation and
the analysis in \cite{Lacquaniti:2009yy}
makes reference the independence
of the metric tensor of $x^5$,
we can claim that the present
work relays more on a cylindricity requirement,
associated to a closed topology, than on a
pure dimensional compactification.
Thus, in \cite{Lacquaniti:2009cr}  the test particle dynamics in a 5D-KK-scenario has
been reformulated
assuming that,  as a consequence of the
unobservable length of the extra dimension, under the Planck scale, test particles are described as localized sources
only in the ordinary 4D-spacetime and delocalized  ones along the
fifth (circular) dimension.
The reformulation \emph{a l$\acute{a}$} Papapetrou of matter and motion is therefore based on the following two assumptions: (1) The metric components and the matter fields do not depend of the extra dimension or
$\partial_5 g_{\ti{AB}}=0$, this assumption is extended also to the 5D--matter tensor or $\partial_5T_{\ti{AB}}=0$
and, (2) The extra dimension has   a circular topology and an unobservable length or
$L_{\ti{(5)}}\equiv\int d^{5}x \sqrt{g_{55}}<10^{-18}\mbox{cm}$.

A 5D-energy momentum tensor $^{\ti{(5)}}\!\mathcal{T}^{\ti{AB}}$ is
associated to the generic 5D-matter distribution governed by the
conservation law $^{\ti{(5)}}\!\nabla_{\ti{A}} \
^{\ti{(5)}}\!\mathcal{T}^{\ti{AB}}=0$,  where $\partial_{5} \
^{\ti{(5)}}\!\mathcal{T}^{\ti{AB}}=0$ (here $^{\ti{(5)}}\!\nabla$
 is the covariant derivative compatible with
the 5D-metric).

Performing a multipole expansion of  $
^{\ti{(5)}}\!\mathcal{T}^{\ti{AB}}$,  centered on a trajectory
$X^{\alpha}$, at the lowest order the procedure gives the equation
of motion for a test particle:
\bea\label{padredimaria}
u^{\mu}\,^{\ti{(4)}}\nabla_{\mu}u^{\nu}=\frac{q}{m}
F^{\nu\rho}u_{\rho}+A(u^{\rho}u^{\nu}-g^{\nu\rho})\frac{\partial_{\rho}\phi}{\phi^3},
\\
\frac{dm}{ds}=-\frac{A}{\phi^3}\frac{d\phi}{ds},
\eea
where $u^{\mu}\equiv dx^{\mu}/ds$,
$g_{\mu \nu}u^{\mu}u^{\nu}=1$,
$F_{\mu\nu}=\partial_{\mu}A_{\nu}-\partial_{\nu}A_{\mu}$ is the
Faraday tensor, being $A_{\mu}$   the U(1) four-vector potential, and $\phi\equiv\sqrt{-g_{55}}$ is the extra KK-scalar field, (see \cite{Lacquaniti:2009cr} and references thereby). Coupling factors to these fields are the charge $q$, coming from the vector current
\be
\api{4}J^{\mu}\equiv \int dV_3 \sqrt{-g_4} \phi
^{\ti{(5)}}\!\mathcal{T}_{5}^{\mu},
\ee
and the scalar charge
\be
A \equiv u^{0} \int dV_3 \sqrt{-g_4} \phi
^{\ti{(5)}}\!\mathcal{T}_{55}.
\ee
Below the definitions for the effective test-particle tensor component follow:
\begin{eqnarray}
\phi \sqrt{-g_4} T^{\mu\nu}&=&\int ds
m \delta^{4}\left(x-X\right) u^{\mu} u^{\nu},
\\\label{qdef}
e k\phi \sqrt{-g_4}
T^{\mu}_{5}&=&\int ds q\delta^{4}\left(x-X\right) u^{\mu}= \sqrt{-g_4}
J^{\mu},
\\\label{Adef}
\phi \sqrt{-g_4} T_{55}&=&\int ds A
\delta^{4}\left(x-X\right).
\end{eqnarray}
The continuity equation $\api{4}\nabla_{\mu}\api{4}J^{\mu}=0$, derived
within the procedure itself, implies that charge $q$ is conserved,
while the requirement $A \equiv 0$, implies that the  mass $m$ is constant. Here we concern on the simplest scenario  where $A=0$, because this choice only allows to recover the case of a constant mass for the particles $m=\mbox{constant}$ \cite{Lacquaniti:2009rh,Lacquaniti:2009wc}.

{For a discussion on the Papapetrou approach to the test particle motion we remind \cite{Lacquaniti:2009yy,Lacquaniti:2009cr,Lacquaniti:2009rq}.}
Therefore, in such a scenario, particles in the GSS spacetimes follow a geodesic equation:
\begin{equation}\label{PapA0}
u^{\mu}\,^{\ti{(4)}}\nabla_{\mu}u^{\nu}=0.
\end{equation}
%
\cite{Lacquaniti:2009cr}, in the following the  notation $^{(4)}$ for four dimensional quantity will be dropped.
The motion of a radially falling particle is regulated by the
$t$-component and $r$-component of the geodesic equation
(\ref{PapA0}), with  $\dot{\theta}=\dot{\varphi}=0$.
%
The velocity  in the $r$-direction is $v_{r}\equiv r'=dr/dt$. For a
particle starting from a point at (spatial) infinity we  have:
\be \frac{dr}{dt}=-\Delta^{\epsilon k-\epsilon/2}
\sqrt{1-\Delta^{\epsilon k}}.
\ee
The radius $\varrho$, at which the radial velocity $v_{r}$ starts to
decreases, is
\be \label{varrho} \varrho=2M \left[1- \left(\frac{2k-1}{
3k-1}\right)^{1/(\epsilon k)}\right]^{-1}\, , \ee
which in the Schwarzschild limit reduces to $\varrho=6M$, otherwise
we have $\rho=\rho(k)<6M$.
The locally measured radial velocity is $v_{r_*}\equiv r'_*=dr^*/dt$
where $dr/dr^*=\sqrt{-g_{tt}/g_{rr}}=\Delta^{\epsilon k-\epsilon/2}$. Therefore we can write:
\be
\frac{dr^*}{dt}=-\sqrt{1-\Delta^{\epsilon k}}.
\ee
The velocities $v_{r}$ and $v_{r^*}$ converge to zero at
(spatial) infinity. The radial velocity $v_{r}$ goes to zero in the
limit $r\rightarrow2M$, and the local measured velocity $v_{r^*}$
goes to $(-1)$ in the limit $r\rightarrow2M$
\cite{Lacquaniti:2009yy,Lacquaniti:2009cr,Lacquaniti:2009rh,Lacquaniti:2009wc}.
Light propagation is described by the geodetic equation
(\ref{PapA0}) with  $d^2s=0$.

This equation, for weak fields,  has been solved in \cite{Kalligas:1994vf},
neglecting the fifth component of the particle velocity, to yield a
hyperbolic orbit. For $k>1$ the defection angle lies in
the range of values which predicted a possible
detectable deviation from the general relativistic values but excludes
null defection and light repulsion.
In particular, gravitational redshift  parameter $z$, at first order
in $r/M$, is
\begin{equation}\label{redred}
z\equiv\frac{\nu_{r}-\nu_{e}}{\nu_{e}}= \epsilon k M
\left(\frac{1}{r_r}-\frac{1}{r_e}\right)
\end{equation}
where $ r_r$ ($\nu_r$) and  $ r_e$ ($\nu_e$)  are the positions
 of the receiver and emitter (frequencies of the received and emitted wave) respectively.
In the Schwarzschild limit this result  is in agreement with the
general relativistic redshift parameter $z_{sch}$ as calculated in
the Schwarzschild geometry; otherwise, for $k>1$ we have
$z=z(k)<z_{sch}$.
\section{Electromagnetic radiation emitted by a radially falling particle.}\label{sec:perturbation}
The charged particle motion and the emitted EM radiation will be
considered as perturbations of the background metric, therefore we
 assume that the EM radiation is regulated by the following
equation
\begin{equation}\label{1801}
\partial_{\nu}\left(\sqrt{-g_4}\phi^3f^{\mu\nu}\right)= 4 \pi\sqrt{-g_4}
J^{\mu},
\end{equation}
where $f^{\mu\nu}$ is the Faraday tensor, see
\cite{Lacquaniti:2009rq}.

The electromagnetic perturbation $f_{\mu\nu}$ is expanded in (4D)-tensor
harmonics $(Y^*_{lm}(\Omega(t)))$. In agreement with the Zerilli
notation \cite{Zerilli:1971wd,Zerilli:1974ai,RuRR} $(\tilde{f}^{\mu\nu})$ denotes the radial
(angle-independent) part     of the Faraday tensor, while the  4D-current $J^{\mu}$
reads
\begin{equation}\label{jmu}
J^{\mu}=\sqrt{4G}\phi T^{\mu}_{5}=\left(\sqrt{-g^4}\right)^{-1}\int
ds q\delta^{4}\left(x-X\right) u^{\mu}\, ,
\end{equation}
where $q$ denotes the charge of the particle, see Eq.\il(\ref{qdef}) and  \cite{Lacquaniti:2009rq,Lacquaniti:2009cr}.

The harmonic coefficients for the
$t$-component and $r$-component of the current source are
respectively:
\begin{eqnarray}
\psi&=&-\frac{q}{r^2}\Delta^{2\epsilon
k-1-3/2\epsilon}\delta(r-R(t)) Y^*_{lm}(\Omega(t)),
\\
\label{ETA}
\eta&=&\frac{q}{r^2}\Delta^{-(\epsilon/2+1)}\frac{dR}{dt}\delta(r-R(t))
Y^*_{lm}(\Omega(t))\, .
 \end{eqnarray}
The independent electric multipole equations are
\be\label{0oOQ} i
\omega\tilde{f}_{01}+\frac{l(l+1)}{r^2}\tilde{f}_{12}\Delta^{2\epsilon
k-1-\epsilon}=-4\pi\eta\Delta^{\epsilon k+\frac{3}{2}\epsilon},
\ee
and
\be\label{1lLIi} i \omega \tilde{f}_{02}\Delta^{\epsilon/2-\epsilon
k}+\partial_{r}\left(\Delta^{\epsilon k
-\epsilon/2}\tilde{f}_{12}\right)+3\tilde{f}_{12}\Delta^{\epsilon
k}\partial_{r}\left(\Delta^{\epsilon/2}\right)=0\,.
\ee
Via the Eqs.~(\ref{0oOQ},\ref{1lLIi}) and  using the homogeneous
Maxwell equation,
\begin{equation}\label{32}
\partial_{r}\tilde{f}_{02}+\imath\omega\tilde{f}_{12}
-\tilde{f}_{01}=0\, ,
\end{equation}
according with the Zerilli's procedure, we find the following
equation for the component $\tilde{f}_{12}$:
\bea
\nonumber
\left[\Delta^{\epsilon(k-1/2)}\left(\Delta^{\epsilon(k-1/2)}
\tilde{f}_{12}\right)_{,r}\right]_{,r} +
\\
\nonumber
3\left[\tilde{f}_{12}
\Delta^{\epsilon(2k-1/2)}\left(\Delta^{-\epsilon/2}\right)_{,r}\right]_{,r}
+
\\ \label{e:e}
\tilde{f}_{12}\left(\!\omega^2-\frac{l(l+1)}{r^2}
\Delta^{\epsilon(2k-1)-1}\!\right)= 4\pi\eta \Delta^{\epsilon
k+3/2\epsilon}.
\eea
Finally  we find, from  Eq.\il(\ref{e:e}) and Eq.\il(\ref{ETA}), the following equation for the function
$\tilde{f}(\omega,r)\equiv \Delta^{\epsilon(k-5/4)} \tilde{f}_{12}$
\begin{equation}\label{questa}
\frac{d^2\tilde{f}(\omega,r)}{dr_{*}^2}
+\left[\omega^2-V_{l}(r,k)\right]\tilde{f}(\omega,r)=S_{l}(\omega,r),
\end{equation}
where $dr/dr_{*}=\Delta^{\epsilon(k-1/2)}$. The source  term reads
\begin{equation}\label{sorx}
S_{l}(\omega,r)= \Delta^{\epsilon(2k-1)} 2 q\sqrt{\frac{1}{2}+l} \
\frac{e^{i\omega t(r)}}{r^2} \Delta^{-(1+\epsilon/2)},
\end{equation}
where  $t(r)$   is a solution of the radial geodesic equation or
\begin{equation}\label{aff}
\frac{dt(r)}{dr}=-\frac{\Delta^{-k\epsilon}}{\sqrt{\Delta^{-\epsilon
}\left(1-\Delta^{\epsilon k}\gamma^{-2}\right)}},
\end{equation}
written as function of $r$ and
$\gamma\equiv\frac{1}{\sqrt{1-v_{\infty}^2}}$, where $v_{\infty}$ is
the (radial)    particle velocity  at spatial infinity.  The
potential $V_{l}(r,k)$ is
%
\begin{equation}
\label{vek}V_{l}(r,k)=\frac{l(l+1)\Delta^{2\epsilon k-\epsilon-1}}
{r^2}+V_{0}(r,k),\end{equation}
where
\begin{equation}\label{golpe2010}
V_{0}(r,k)\equiv\frac{3\epsilon }{4}\frac{M^2\Delta^{2\epsilon
k-\epsilon-2}} {r^4}\left[4\left(\epsilon
k+1-\frac{r}{M}\right)+\epsilon\right].
\end{equation}
{(see\cite{Zerilli:1971wd,Zerilli:1974ai,RuRR} for current use of the notation)}.
In the  Schwarzschild limit $V_{0}(r,k)=0$ and
\begin{equation}\label{golpe93}
V_{l}(r,k)\equiv V_{l}(r)=\frac{l (1+l)\Delta}{r^2}
\end{equation}
We approached the Sturm-Liouville problem for the function
$\tilde{f}(\omega,r)$ in Eq.\ref{questa} using  the Green's
functions method. We first solved the  associated homogeneous
 equation (obtained  from Eq.\ref{questa} imposing
$S_{l}(\omega,r)=0$).  We used the solutions of this equation   to
build the Green's function $G(r,r')$ of the problem.
%
Since we are interested in  the EM spectra as seen by an observer
located at (spatial) infinity  we can focus on the outgoing
component only of the EM radiation. We require that the outgoing radiation
is purely oscillating  at infinity.
%
The  energy spectra at (spatial) infinity
reads
\begin{equation}\label{inifm}
\frac{dE}{d\omega}=\sum_{l}\frac{dE_{l}}{d\omega}=\sum_{l}\frac{l(l+1)}{2
\pi}\left|A^{out}_{l}(\omega)\right|^{2},\end{equation}
where $\omega\geq0$ and  where $A^{out}_{l}$ is a function of
$\omega$ only.
The homogeneous equation is a ``Schrodinger-like'' equation for the
function $\tilde{f}(\omega,r^*(r))$ with a potential
$V_{l}(r^*(r),k)$.  Asymptotically, for large value of $r$, the
potential $V_{l}(r,k)$ decreases to zero.

The  values of $V_{l}(R,k) M^2$ in $R\equiv(2+10^{-a})M$ for
different values of  $a$, and $k$ are listed in Table
\ref{Table:G9}.
\begin{table}[hbpt]
\begin{center}
\begin{tabular}{lcccccr}\noalign{\smallskip}
\hline
 a &-6& -10& -50& -80\\
 \hline
k=5& 0.0098& 0.01367& 0.3777& 4.553\\
 k=10& 0.0016&
0.0018& 0.0038& 0.0067\\
\hline
\end{tabular}
\end{center}
\caption[font={footnotesize,it}]{The function
$V_{l}(R,k) M^2$ calculated at $l=1$, and $R\equiv(2+10^{-a})M$, for
different values of $k$ and $a=\{-6,-10,-50,-80\}$.}
\label{Table:G9}\end{table}
\subsection{1}\label{quasi}
At first we investigate the problem in a quasi-Schwarzschild regime.
Therefore we expand Eq.\ref{questa} at first order in $1/k$ around $1/k=0$, neglecting  higher order terms. The potential
$V_{l}(r,k)$ vanishes as $r$ approaches $2M$. We look for a solution
$f_{l}$, of the homogeneous equation (\ref{questa}) with the following asymptotical behavior
%
\begin{eqnarray} \label{aficionados} f_{l}&\rightarrow&
e^{-i \omega r_*}\quad\mbox{as}\quad r_{*}\rightarrow - \infty\\
\label{aficionados9} f_{l}&\rightarrow& B_l(\omega)e^{+i
\omega r_*}+C_l(\omega)e^{-i \omega r_*}\,\mbox{as}\,
r_{*}\rightarrow+\infty
\end{eqnarray}
%
where, because we are taking the limit of large $k$, $r^{*}=r+2M
Log[r/(2M)-1]$ and   $B_l(\omega)$ and $C_l(\omega)$ are functions
of the Fourier frequency $\omega$, (see for example\cite{Zerilli:1971wd,Zerilli:1974ai,RuRR,Cardoso:2003cn}).
%

\begin{figure*}
\resizebox{1\textwidth}{!}{%
\includegraphics[scale=.71]{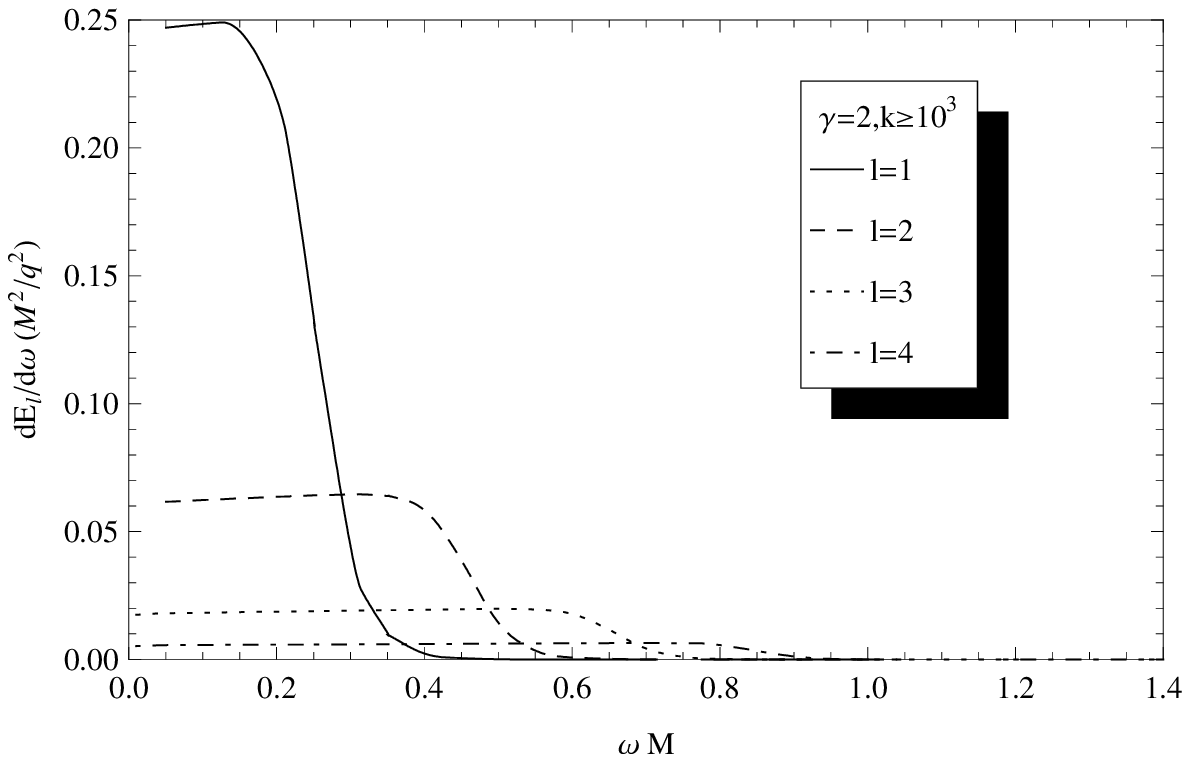}\\
\includegraphics[scale=.71]{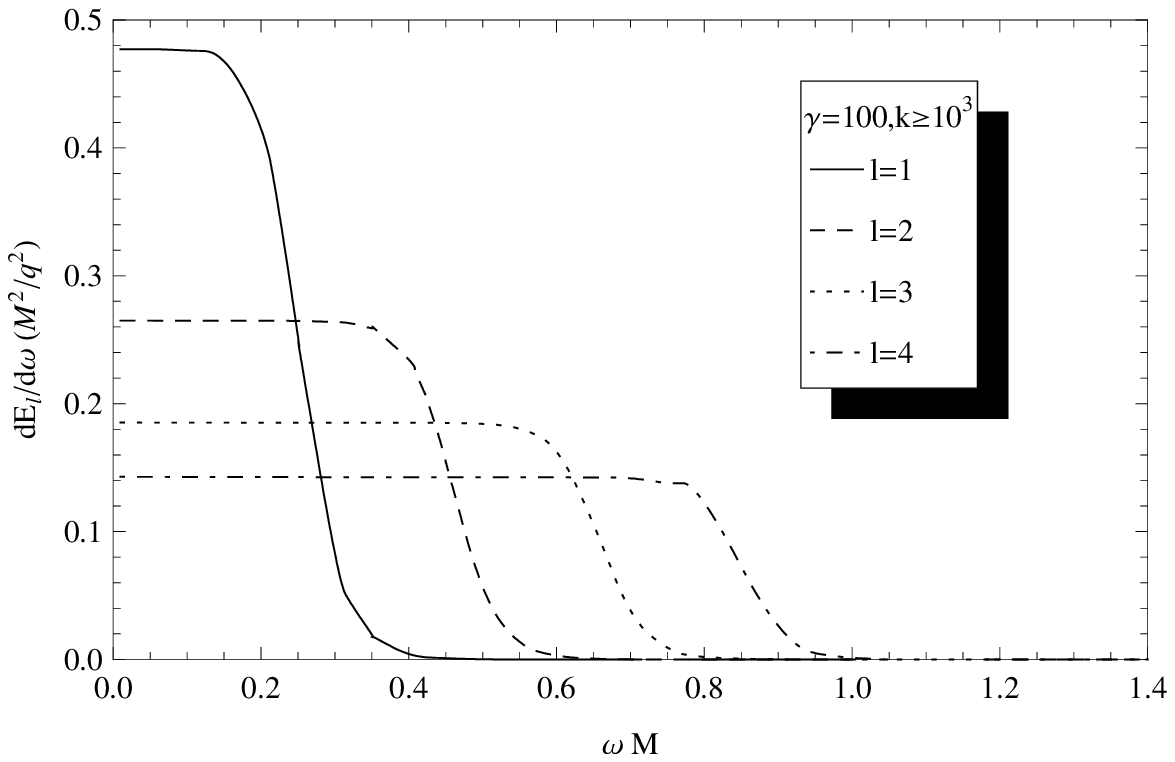}
}\caption{EM-energy spectra $(k=1000)$ for the l-multipole, for
particle falling from infinity into a
GSS-background.}\label{fig:acqua}
\end{figure*}

The problem was approached using numerical integration. Adopting a
Runge-Kutta method, we first integrated the homogeneous equation for
$k=1000$ to obtain an evaluation of the function $f_l$ with the
required boundary condition (\ref{aficionados}). Numerical
integration started at $r=M(2+10^{-6})$. It is stopped at large
values of $r$ as well as the integral in (\ref{ujm}) converges.
Boundary condition (\ref{aficionados9}) is then used to recover a
value of $C_l$. Once the values of the parameters $(\gamma,l)$  are
fixed, the integration has been performed for many fixed values of
$\omega$. The spectra profile,  Fig.\ref{fig:acqua}, coincides with
that emitted  in the Schwarzschild background \cite{Cardoso:2003cn}.
\subsection{2}
\begin{figure}
\resizebox{.51\textwidth}{!}{%
\includegraphics[scale=.71]{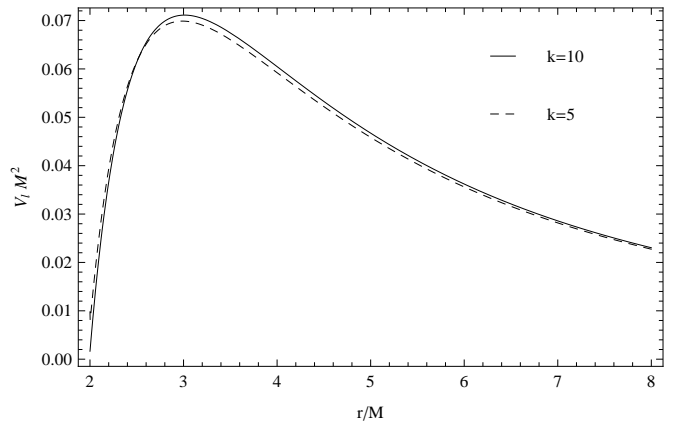}
}
\caption{The potential $V_{l}(r,k)$ for   $l=1$ and $k=10$ (black
line), $k=5$ (dashed line),  $k=2.2$ (dotdashed line).} \label{Fig:Marpra}
\end{figure}

\begin{figure*}
\resizebox{1\textwidth}{!}{%
\includegraphics[scale=.51]{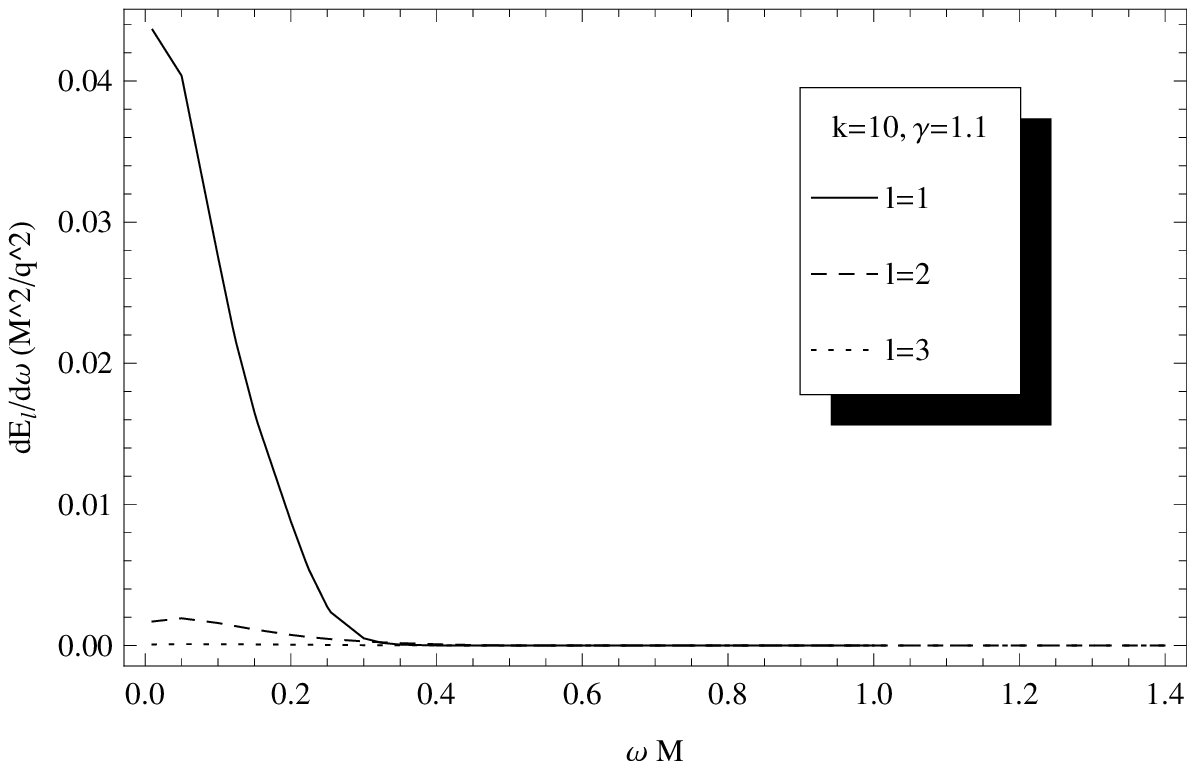}
\includegraphics[scale=.51]{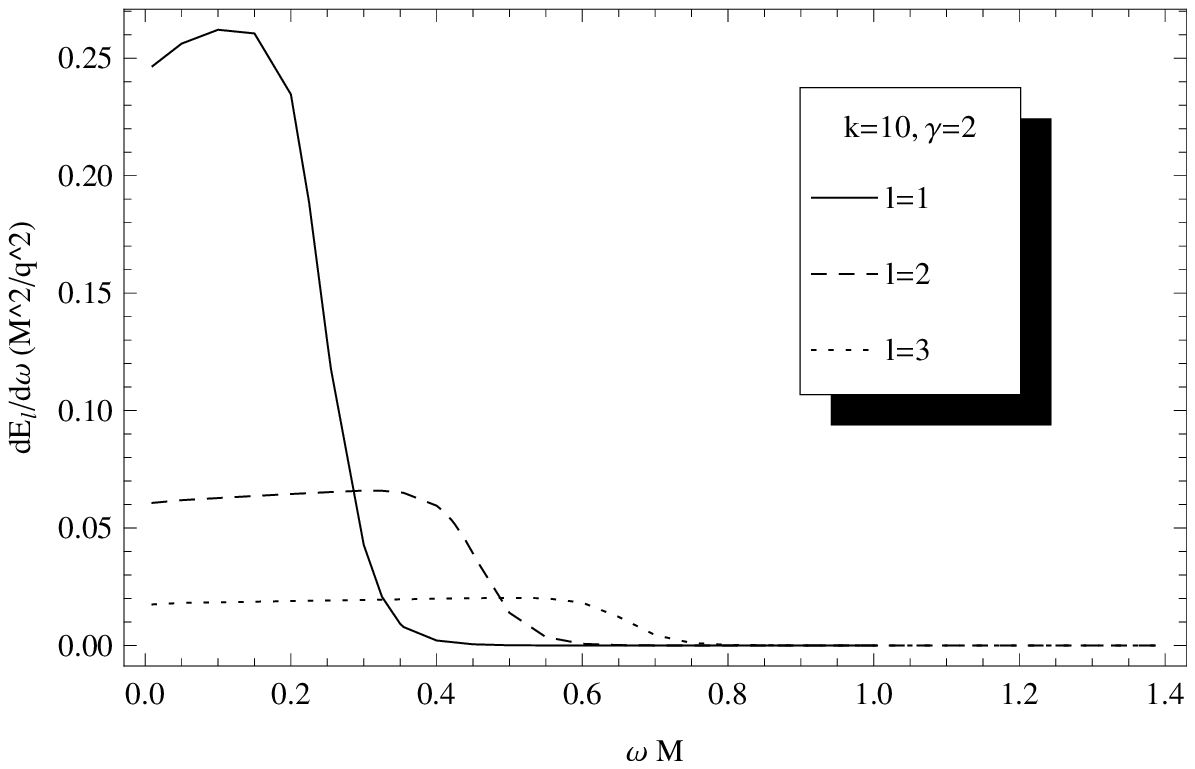}
\includegraphics[scale=.51]{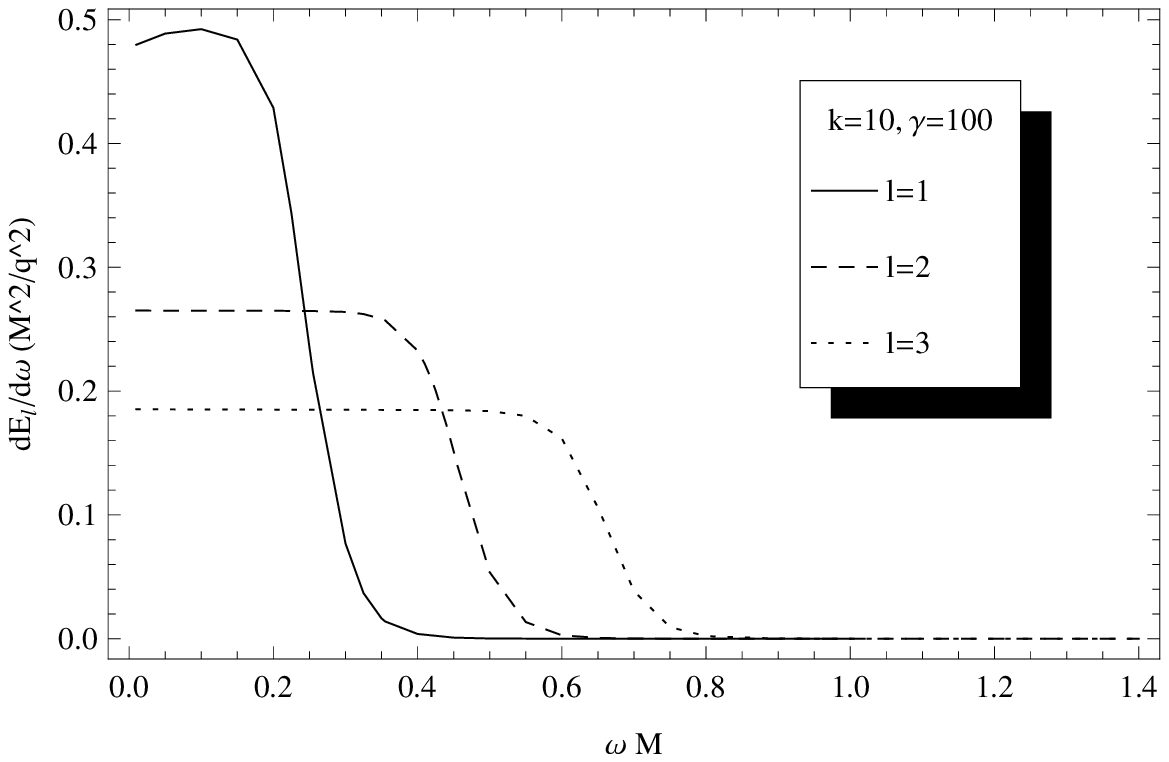}
}
\caption[font={footnotesize,it}]{\footnotesize{Electromagnetic
energy spectra    for particle falling into a GSS-background with
$k=10$ is plotted for different l-multipoles and different values of
$\gamma$. }
}\label{Dulcineak10}
\end{figure*}

\begin{figure*}
\resizebox{1\textwidth}{!}{%
\includegraphics[scale=.51]{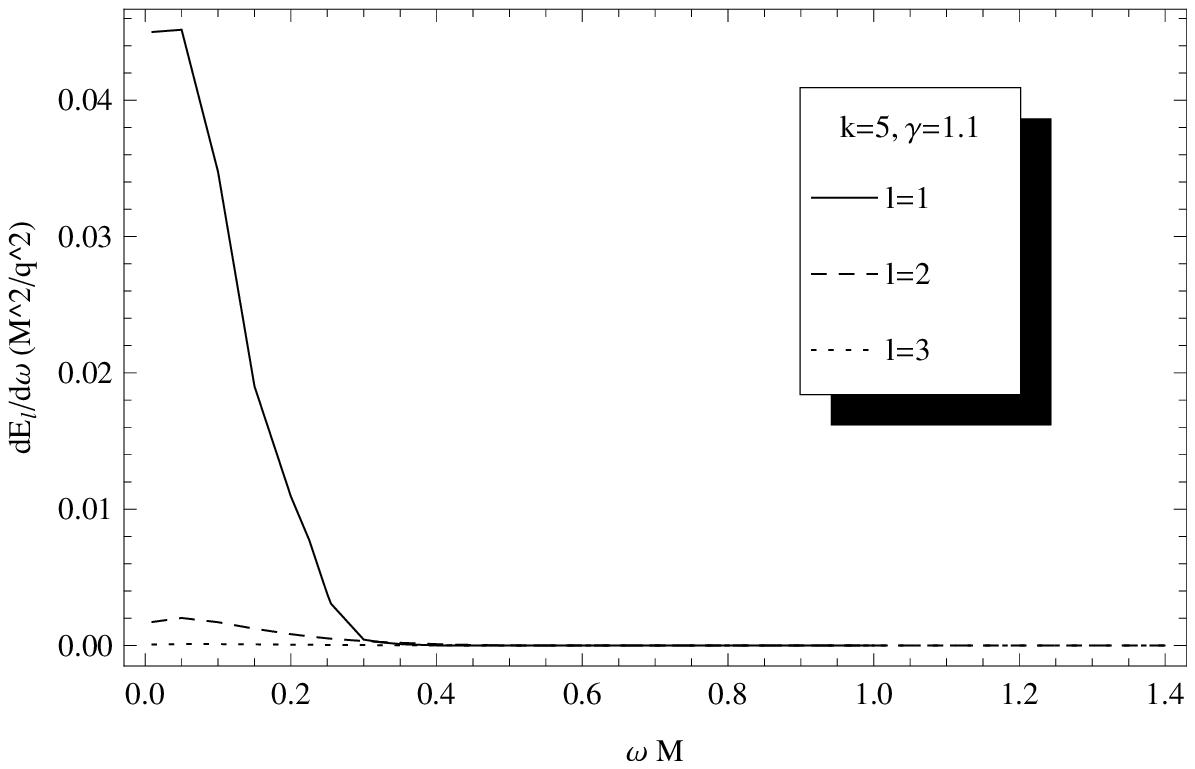}
\includegraphics[scale=.51]{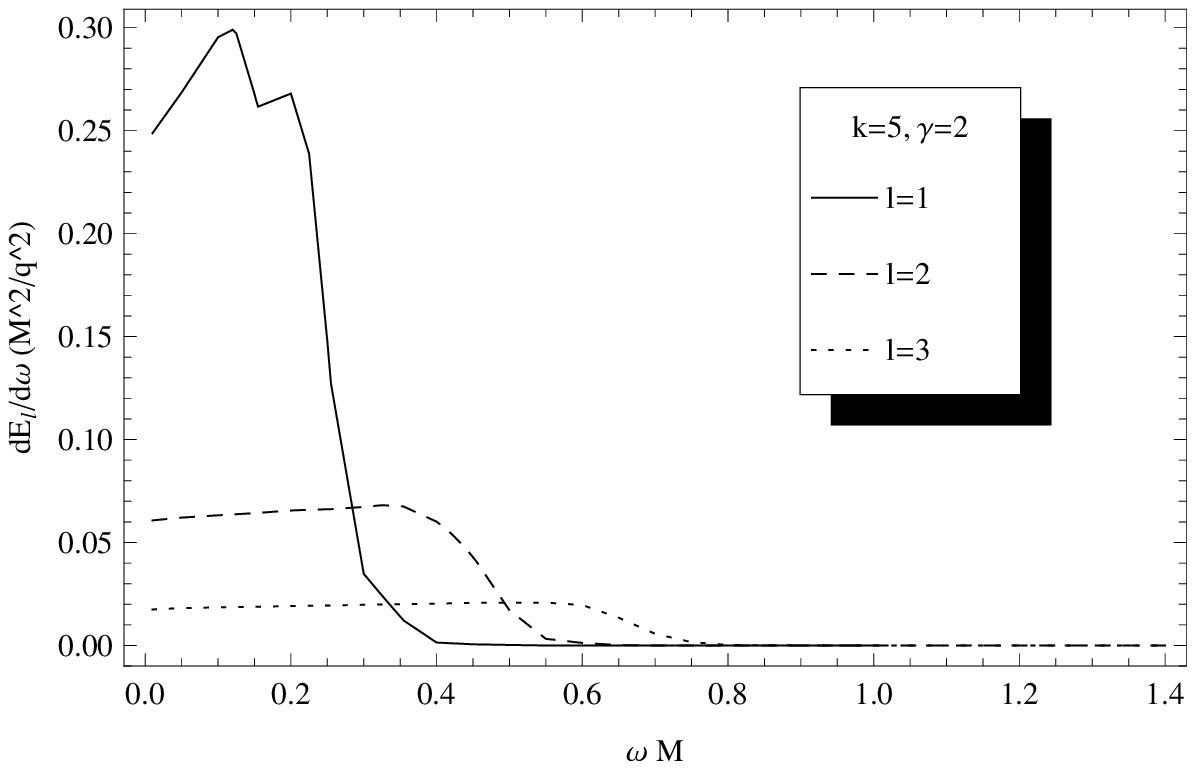}
\includegraphics[scale=.51]{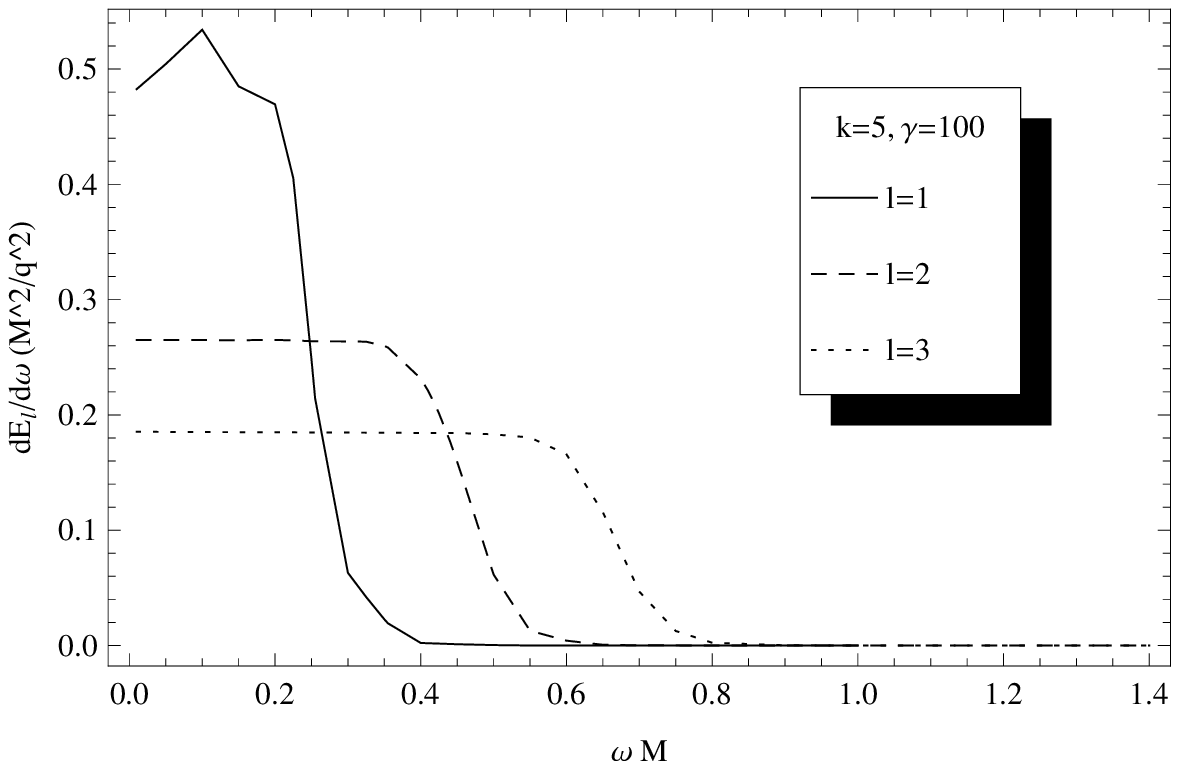}
}\caption[font={footnotesize,it}]{\footnotesize{Electromagnetic
energy spectra    for particle falling into a GSS-background with
$k=5$ is plotted for different l-multipoles and different values of
$\gamma$.} }\label{Dulcineakl}
\end{figure*}

We integrated  Eq.\ref{questa}  numerically
for $r>R=M(2+10^{-6})$ and for $k={10,5}$. The potential is plotted
in Fig. \ref{Fig:Marpra}.
The homogeneous equation is solved imposing that the solution is ingoing
at $R$ and a combination of ``incident'' and ``transmitted'' waves
at spatial infinity.
The $\omega$-function $A^{out}_{l}$ is therefore the  integral
\begin{eqnarray}  \label{ujm}
A^{out}_{l}(\omega)&=& \frac{1}{2i\omega
C_{l}(\omega)}\int_{R}^{\infty}f_{l} \hat{S}_{l}dr,
\end{eqnarray}
where
\begin{equation}\label{sorb}
\hat{S}_{l}(\omega,r)=2q\sqrt{\frac{1}{2}+l}  \frac{e^{i\omega
t(r)}}{r^2}\Delta^{\epsilon(k-1)-1}.
\end{equation}
Spectra profiles are plotted in Fig.~\ref{Dulcineak10},\ref{Dulcineakl}.
%

A general feature of the emitted spectra for $\gamma\geq2$ for the
Schwarzschild's case is to grow up to a critical value
(corresponding to a critical frequency $\omega$ for a fixed
multipole) and then rapidly flow down to zero (see  \cite{Cardoso:2003cn} and also Fig.~\ref{fig:acqua} in this section).
%
For the GSS case  we found that, for a fixed $\gamma$, the spectra
profile coincides with that of Schwarzschild's case as well as $k$
is sufficiently large ($k>5$). However there is an increase of the energy emitted
rate  per frequency $dE/d\omega$, at fixed value of $\omega$ and  fixed multipole $l$. There are significant discrepancies with the
Schwarzschild's case for $k\leq5$ where, for low frequencies, a peak in the spectra profile
appears  also at large $\gamma$.
\section{Conclusions}\label{sec:concldisc}
In principle the relic of an extra dimension could be
recognized in the Universe as a deformation of the EM-emission
spectra from compact astrophysical sources compared with that
allowed by general relativistic calculations. The spectral analysis
could provide therefore  a suitable  test to constrain the
multidimensional theories. We argue in particular that the comparison with the observed data on the spectral emission by collapsing compact object, that could lead at last in principle to the formation of a black hole  {could get light on the nature of the singularity and the possible presence of the small extra  dimension }, for a discussion about the possible constraint of the KK-theory by comparing with strong emission astrophysical processes see for example \cite{Matsuno:2011ca,Bonnevier:2011km,Becar:2011fc,Inte,Cardoso:2003cn,lw,ww}.

{On the other side, a
non trivial question for any theories dealing with one or more  extra-dimensions, arises on the non--observation of such an
extra-space. The most natural answer to this phenomenological
question consists in the assumption of  the existence of a  closed extra-space,
having a volume much smaller that the smallest observed scale
in the present day experiment of high-energy physics, say
$\mathcal{O}(10^{-18}\rm{cm})$.
In the standard Kaluza-Klein approach, the necessity
to reproduce the proper value of the fundamental coupling
constants (in five dimension the fine structure constant)
implies that the extra dimensions be compactified to very small
values, two orders of magnitude above the Planck scale.
Therefore, the request for a Planckian-like compactification
scale can not be completely escaped and we have to deal with
some of the fundamental paradigms proposed for the Kaluza-Klein
scenario \cite{Buchbinder:1989cd,Overduin:1998pn}.
We point out that
the condition of compactification, introduced by Klein, consists in the assumptions that the extra dimension is small and has a closed topology, form this hypothesis we get to a periodicity  an the possibility to perform a  Fourier decompositions, see for example\cite{Overduin:1998pn}.
On the other hand, the cylindricity hypothesis assumes
metric components  do not depend on
the fifth coordinate or
$\partial_5 g_{\ti{AB}}=0$.
In this work we have considered a class of static solutions of
5D-KK-equations independent by the fifth extra-dimension (GSS).
For the observer at radial infinity, the extra-dimension is
not observable because it is on a steady Planckian-like scale,
and for the observers closer to the star surface its dimension
increases according to the increasing behavior of the scale
factor associated to the extra-dimension, namely the scalar
field $\phi(r)$. The scale factor varies only for few orders of
magnitude from infinity to the star surface, so that the
phenomenology of the Kaluza-Klein picture is not affected by
features of our model. Indeed, it is just the different value
that the scale factor (the scalar field) takes near the star,
that is responsible for the observability of a deviation in the
spectrum of the emitted radiation by the infalling particle,
with respect to the ordinary four-dimensional case. }

The closer is the surface of the star to the radial value corresponding in
vacuum to the naked singularity, the stronger will be the
modification that the emitted spectrum is expected to outline.
In fact, it is just the variation of some order of magnitude in
the compactification scale of the fifth dimension that is
observed at radial infinity as a deformation of the spectrum,
since if the scalar field would be equal to unity everywhere,
the solution would reduce to an ordinary Schwarzschild profile
and the morphology of the emission would be unaltered.

We studied the charged particle motion  and the EM-radiation emitted
by 4D-pointlike charged particle freely falling in the selected
 backgrounds. We have considered the particles and the fields
as perturbations of the background metric. The EM-emission spectra,
as observed at infinity, is found.

The spectra profiles, in the GSS metrics  here analyzed,  are in
good agreement with the emitted spectra calculated in the
Schwarzschild case as long as $k$
is sufficiently large ($k>5$).  {Therefore the effects of the presence of an extradimension in the framework of the KK-theory should be evident, from the point of view of the EM emission spectra only in those spacetime characterized by $k<5$. }

For a  fixed value of $\omega$ and a fixed multipole $l$ an increase of the energy emitted
rate  per frequency $dE/d\omega$ is noted. Significant discrepancies respect to   the
Schwarzschild case appear for $k\leq5$ where, for low frequencies, a peak in the spectra profile
appears  also at large $\gamma$.

The development of the present analysis suggests to explore the EM emission by charged particle in the 5D-vacuum spacetime. This analysis should be interesting for many reasons and in particular  for the exploration of the naked singularity features of these vacuum solutions.
The main problem we could recognize in the treatment of these configurations from the point of view of the EM emission spectra is faced upon the effective potential in Eq. \ref{questa}.
The different physical situations are clearly distinguished by  the boundary conditions to be imposed on the Cauchy problem,  (see also  \cite{Dotti:2008ta}).

In the vacuum spacetime the effective potential, for each finite value of the $k$-parameter  explodes on the ring  $r=2M$, meanwhile in the Schwarzschild case, reached on the 4D-spacetime section and in the limit of large $k$, it is zero on the horizon $2M$. The boundary condition, and therefore the spectra profile obviously remark this different nature of the circle $r=2M$. In the Schwarzschild limit case in the vacuum GSS, see Section (\ref{quasi}), we impose the ingoing  boundary condition a the horizon, in the vacuum GSS background with $k$ finite we attend to impose on the infinite wall of potential on the  singularity $r=2M$ the wave function $f_l$ annihilates.

A remarkable feature of this process is therefore the contrast between the two vacuum cases not resolvable a part in the limit $k\rightarrow\infty$. On the other hand if  we compare the dynamics of timelike particles and even photons  this breakdowns of these two physical situations does not appear.
We finally observe that, this idea that we
can measure at a very large distance from
a compact object,
{features accounting for its different
status of the extra dimension scale, $\phi$,  in the five-dimensional
Kaluza-Klein scenario, is the just the core of the
idea we are proposing by the present calculation}.

Thus, the analysis of the motions in the vacuum spacetime  in \cite{Lacquaniti:2009wc,Lacquaniti:2009rh} within the
Papapetrou approach should be  therefore completed by  the  analysis  of the emission processes induced by charged particles in the (vacuum) GSS background.
The comparison in  particular of the electromagnetic emission spectra  due to a freely falling charged test particle, a   in the GSS and in Schwarzschild background could get light on physics around the naked  singular ring $r=2M$ (\cite{Virbhadra:2007kw,Virbhadra:2002ju} and \cite{lw,ww}).
The solution of the equations governing the emission process should strongly depend on the boundary conditions imposed on the equations. The comparison between the electromagnetic emission by radially falling charged particles in  the GSS and stellar case could be strongly discriminant.

\begin{acknowledgement} We wish to acknowledge  M. Lattanzi for
help in what concerns the numerical integration. This work has been developed in the framework of the CGW Collaboration (www.cgwcollaboration.it).
One of us (DP) gratefully acknowledges financial support from the Angelo Della Riccia Foundation.
\end{acknowledgement}

\bibliographystyle{elsarticle-num}

\begin{thebibliography}{99}

\bibitem{Matsuno:2011ca}
  K.~Matsuno and K.~Umetsu,
  Phys.\ Rev.\  D {\bf 83} (2011) 064016.


\bibitem{Rizzo:1999qb}
  T.~G.~Rizzo,
  AIP Conf.\ Proc.\  {\bf 530} (2000) 290.
%

\bibitem{Eingorn:2010wi}
  M.~Eingorn and A.~Zhuk,
  Phys.\ Rev.\  D {\bf 83} (2011) 044005.



\bibitem{Moutsopoulos:2011ez}
  G.~Moutsopoulos and P.~Ritter,
  arXiv:1103.0152.


\bibitem{Stelea:2009ur}
  C.~Stelea, K.~Schleich and D.~Witt,
  Phys.\ Rev.\  D {\bf 83} (2011) 084037.

\bibitem{Bonnevier:2011km}
  J.~Bonnevier, H.~Melbeus, A.~Merle and T.~Ohlsson,
  arXiv:1104.1430.


\bibitem{Okawa:2011fv}
  H.~Okawa, K.~i.~Nakao and M.~Shibata,
  arXiv:1105.3331 [gr-qc].

\bibitem{Tomizawa:2011mc}
  S.~Tomizawa and H.~Ishihara,
  arXiv:1104.1468.

 \bibitem{Moon:2011sz}
  T.~Moon, Y.~S.~Myung and E.~J.~Son,
  arXiv:1104.1908.


\bibitem{Becar:2011fc}
  R.~Becar and P.~A.~Gonzalez,
  arXiv:1104.0356.

\bibitem{Yamada:2011br}
  Y.~Yamada and H.~a.~Shinkai,
  Phys.\ Rev.\  D {\bf 83} (2011) 064006.

\bibitem{Inte}
Q. Jiang, S.Yang, H. Li, Int. Journ. of Theoretical Physics, Vol. 45, No. 9, (2006).





\bibitem{Librogra}
\emph{Modern Kaluza-Klein Theories}, edited by T. Applequist, A. Chodos,
and P.G.O. Freund, Addison-Welsey, Menlo Park, 1987.

\bibitem{PSWesson}
 P. S. Wesson, \emph{Space-Time-Matter, Modern
Kaluza-Klein Theory}, World Scientific 1999.

\bibitem{P.D.B.Collins}
P. D. B. Collins, A. D.
Martin and E. J. Squires, \emph{Particle Physics and Cosmology}, Wiley
1989.

\bibitem{Cianfrani:2009wj}
  F.~Cianfrani and G.~Montani,
  arXiv:0904.0574 [gr-qc].

\bibitem{Overduin:1998pn}
  J.~M.~Overduin and P.~S.~Wesson,
  Phys.\ Rept.\  {\bf 283} (1997) 303.

%
\bibitem{Wesson:1999nq}
  P.~S.~Wesson,
{\it  Singapore, Singapore: World Scientific (1999) 209 p}.

\bibitem{Kokkotas:1999bd}
  K.~D.~Kokkotas and B.~G.~Schmidt,
  Living Rev.\ Rel.\  {\bf 2} (1999) 2.
\bibitem{Lacquaniti:2009wc}
  V.~Lacquaniti, G.~Montani, D.~Pugliese,
 Gen.\ Rel.\ Grav.\
 DOI:10.1007/s10714-010-1007-3.


\bibitem{Xu:2007dc}
  P.~Xu and Y.~g.~Ma,
  Phys.\ Lett.\  B {\bf 656} (2007) 165.

\bibitem{Overduin:2000gr}
  J.~M.~Overduin,
  Phys.\ Rev.\  D {\bf 62} (2000) 102001.
\bibitem{PoncedeLeon:2006xs}
  J.~Ponce de Leon,
  Int.\ J.\ Mod.\ Phys.\  D {\bf 17} (2008) 237.

\bibitem{PoncedeLeon:2007bm}
  J.~Ponce de Leon,
  Int.\ J.\ Mod.\ Phys.\  D {\bf 18} (2009) 251.




\bibitem{Lacquaniti:2009rq}
  V.~Lacquaniti and G.~Montani,
  Mod.\ Phys.\ Lett.\  A {\bf 24} (2009) 1565.

  \bibitem{KKStars}
   D.~Pugliese and G.~Montani,
  Eur.\ Phys.\ J.\  C {\bf 71} (2011) 1638.

\bibitem{Capozziello:2011nr}
  S.~Capozziello, M.~De Laurentis, S.~D.~Odintsov and A.~Stabile,
  Phys.\ Rev.\  D {\bf 83} (2011) 064004.



\bibitem{Casadio:2003iv}
  R.~Casadio, S.~Fabi and B.~Harms,
  Phys.\ Rev.\  D {\bf 70} (2004) 044026.


\bibitem{Lacquaniti:2009yy}
  V.~Lacquaniti and G.~Montani,
  Int.\ J.\ Mod.\ Phys.\     D {\bf 18} (2009) 929.

\bibitem{Lacquaniti:2009cr}
  V.~Lacquaniti and G.~Montani,
  arXiv:0906.2231 [gr-qc].

\bibitem{Lacquaniti:2009rh}
  V.~Lacquaniti, G.~Montani, D.~Pugliese
  arXiv:0911.4168 [gr-qc],  in Proceedings of
MG12, Marcel Grossman Meetin, Paris, 2009 (World
Scientific, Singapore, 2011).

\bibitem{Cianfrani:2005em}
  F.~Cianfrani, A.~Marrocco and G.~Montani,
  Int.\ J.\ Mod.\ Phys.\  D {\bf 14} (2005) 1195.

\bibitem{Cianfrani:2006sc}
  F.~Cianfrani and G.~Montani,
  Int.\ J.\ Mod.\ Phys.\  D {\bf 17} (2008) 785.




\bibitem{Montani:2009wg}
  G.~Montani, N.~Carlevaro, F.~Cianfrani and V.~Lacquaniti,
  arXiv:0904.0568 [gr-qc].

\bibitem{papapetrou}
A. Papapetrou, \emph{Proc. Phys. Soc.} \textbf{64}, 57 (1951).




\bibitem{Kalligas:1994vf}
  D.~Kalligas, P.~S.~Wesson and C.~W.~F.~Everitt,
  Astrophys.\ J.\  {\bf 439} (1994) 548.
\bibitem{Zerilli:1971wd}
  F.~J.~Zerilli,
  Phys.\ Rev.\   D {\bf 2} (1970) 2141.
\bibitem{Zerilli:1974ai}
  F.~J.~Zerilli,
Phys.\ Rev.\   D {\bf 9} (1974) 860.

 \bibitem{RuRR} R.
Ruffini \emph{On the Energetics of Black Holes, Le Astres Occlus}
Les Houches.


\bibitem{Cardoso:2003cn}
  V.~Cardoso, J.~P.~S.~Lemos and S.~Yoshida,
  Phys.\ Rev.\   D {\bf 68} (2003) 084011.

\bibitem{Virbhadra:2007kw}
  K.~S.~Virbhadra and C.~R.~Keeton,
  Phys.\ Rev.\  D {\bf 77} (2008) 124014.



\bibitem{Virbhadra:2002ju}
  K.~S.~Virbhadra and G.~F.~R.~Ellis,
  Phys.\ Rev.\  D {\bf 65} (2002) 103004.


 \bibitem{lw}
H. Liu and P. S. Wesson,
J. Math. Phys. 42, 4963 (2001).





\bibitem{ww}
M. Cassé, J. Paul, G. Bertone,  G. Sigl
Phys. Rev. Lett. 2004 Mar 19;92(11):111102.



\bibitem{Dotti:2008ta}
  G.~Dotti and R.~J.~Gleiser,
  Class.\ Quant.\ Grav.\  {\bf 26} (2009) 215002.









\bibitem{Buchbinder:1989cd}
  I.~L.~Buchbinder and S.~D.~Odintsov,
  Fortsch.\ Phys.\  {\bf 37} (1989) 225.






 \end{thebibliography}

\bibliographystyle{model1-num-names}

\end{document}